%% file: 0Main.tex
\documentclass[conference]{IEEEtran}

\usepackage{caption}
\usepackage{{booktabs}}
\usepackage{url}
\usepackage{cite}
\usepackage{graphicx} 
\usepackage{tabularx}
\usepackage{multirow}
\newcolumntype{Y}{>{\centering\arraybackslash}X}
\usepackage{array}
\usepackage{rotating}
\usepackage{float}
\usepackage{pgfplots}
\usepgfplotslibrary{statistics}
\usepackage{csvsimple}
\pgfplotsset{compat=1.18}
\usepackage[spanish, english]{babel}
\usepackage{pgf-pie}
\usepackage{enumitem}
\usepackage{tikz}
\usetikzlibrary{shapes,backgrounds,positioning,patterns}
\pgfplotsset{
    discard if not/.style 2 args={
        /pgfplots/boxplot/data filter/.code={
            \edef\tempa{\thisrow{#1}}
            \edef\tempb{#2}
            \ifx\tempa\tempb
            \else
                
            \fi
        }
    }
}

\begin{document}
%
\title{Exploring Content Concealment in Email}



\author{\IEEEauthorblockN{Lucas Betts}
\IEEEauthorblockA{School of Computer Science,\\
The University of Auckland,\\
Auckland, New Zealand.\\
lucas.betts@auckland.ac.nz}
\and
\IEEEauthorblockN{Robert Biddle}
\IEEEauthorblockA{School of Computer Science,\\
Carleton University,\\
Ottawa, Canada.\\
robert.biddle@carleton.ca}
\and
\IEEEauthorblockN{Danielle Lottridge}
\IEEEauthorblockA{School of Computer Science,\\
The University of Auckland,\\
Auckland, New Zealand.\\
d.lottridge@auckland.ac.nz}
\and
\IEEEauthorblockN{Giovanni Russello}
\IEEEauthorblockA{School of Computer Science,\\
The University of Auckland,\\
Auckland, New Zealand.\\
g.russello@auckland.ac.nz}
}

%


\newcommand\myworries[1]{\textcolor{red}{#1}}
\newcommand\tocite[1]{\textcolor{blue}{#1}}

\maketitle

\begin{abstract}
\input{1Abstract}
\end{abstract}


%
\IEEEpeerreviewmaketitle

 

\section{Introduction}
\input{2Introduction}

\section{Related Work}
\input{3Literature}

\section{Research Gap}
\input{4ResearchGap}

\section{Methodology}
\input{5Methodology}

\section{Results}
\input{6Results}

\section{Discussion}
\input{7Discussion}

\section{Conclusions}
\input{8Conclusion}






\input{0main.bbl}

\newpage
\onecolumn
\section{Appendix}
\input{9Appendix}

\end{document}

%% file: 1Abstract.tex
The never-ending barrage of malicious emails, such as spam and phishing, is of constant concern for users, who rely on countermeasures such as email filters to keep the intended recipient safe. 
Modern email filters, one of our few defence mechanisms against malicious emails, are often circumvented by sophisticated attackers.
This study focuses on how attackers exploit HTML and CSS in emails to conceal arbitrary content, allowing for multiple permutations of a malicious email, some of which may evade detection by email filters. 
This concealed content remains undetected by the recipient, presenting a serious security risk.
Our research involved developing and applying an email sampling and analysis procedure to a large-scale dataset of unsolicited emails.
We then identify the sub-types of concealment attackers use to conceal content and the HTML and CSS tricks employed.

%% file: 2Introduction.tex
Email is one of the most essential methods of communication, with over 340 billion emails sent daily \cite{noauthor_email_nodate}. This ubiquity makes email a prime target for malicious actors who exploit it to defraud individuals and organizations, necessitating robust email security measures. 
Despite extensive research, attackers continuously evolve their methods, outpacing current countermeasures \cite{oliveira_empirical_2019, ferrara_history_2019, noauthor_apwg_2024}.
One specific threat is content concealment in emails, where attackers insert hidden content that is visible to email filters but invisible to recipients, avoiding detection by email filters without modifying the email for the recipient.\par\smallskip

The goal of an email filter is to classify all malicious emails without falsely identifying legitimate emails as malicious. Attackers aim to deceive these filters into marking their emails as legitimate. Email filters utilize various features and methods, including pattern matching and heuristic filters, where matching patterns adjust the likelihood of an email being flagged as malicious. Although commercial filters often keep their rules secret, open-source examples like SpamAssassin provide insight into potential rules \cite{noauthor_spamassassin_nodate}. Additionally, filters often use statistical and machine learning-based methods, where rules are not fixed, and decisions evolve with new data, resulting in highly accurate email filtering with low false positive and negative rates\cite{chaudhry_phishing_2016}.\par\smallskip

Commercial email filters are significant barriers for attackers aiming to reach recipients’ inboxes. While many attackers rely on volume, motivated attackers can pilot-test their emails against these filters by sending permutations to monitored inboxes. This method allows attackers to find versions that pass through the filters. However, this approach requires modifications that may alert recipients to the email’s malicious intent, similar to early Bayesian poisoning attacks against statistical text filters \cite{lowd_good_nodate, wittel_attacking_nodate}.\par\smallskip

Emails can contain Hypertext Markup Language (HTML), Cascading Style Sheets (CSS), and other attachments. Attackers can utilise intended features of these languages to conceal arbitrary content into the body of an email, allowing for permutations of an email to be created without modifying the rendered version of the email\cite{bergholz_detecting_nodate, moens_identifying_2010, fumera_spam_nodate, janez-martino_review_2023}. The result is attackers can now use these insertions to attempt to bypass email filtering systems without the recipient being alerted.\par\smallskip

In this study, we introduce and implement a novel email sorting system designed to uncover hidden content in malicious emails within a comprehensive dataset. Our methodology then involved a meticulous manual examination of emails, leading to the categorization of three primary sub-types of content concealment utilised by attackers. Furthermore, we conducted an in-depth analysis of the specific CSS tricks used by attackers in each of the three key sub-types of concealment. This exploration firstly highlights the CSS tricks used by attackers but also prompts a need for discussion on the restriction of these CSS features to enhance email security. The significant contributions of our work lie involved developing and applying an email sampling and analysis procedure, the detailed categorization of concealment sub-types, and the identification of the key CSS tricks utilised for these concealment sub-types. Collectively, this work offers a clearer understanding of attacker capabilities and helps pave the way for more robust defences against email-based threats, thereby addressing a critical gap in current email security measures.

%% file: 3Literature.tex
Terminology such as ``hidden text salting'' was used by many of the early authors such as Bergholz and Moens \cite{bergholz_detecting_nodate, moens_identifying_2010}, yet was expanded to cover more concepts such as ASCII murals or visible character insertions by later authors, for example, \foreignlanguage{spanish}{Jáñez‑Martino} \cite{janez-martino_classication_2020, janez-martino_review_2023}. Due to this shift in terminology over the years, we developed a new phrase, ``content concealment,'' to refer specifically to text in the body of an email intended not to be visible to the potential recipient of an email.\par\smallskip

Wittel and Wu \cite{wittel_attacking_nodate} are one of the earliest studies addressing the implication of content concealment techniques used by spammers in 2004. They identify the impact of content concealment and hidden text salting on the efficacy of spam filters.\par\smallskip

Further research was conducted by Bergholz et al. in 2008 \cite{bergholz_detecting_nodate} using a combination of optical character recognition (OCR) and machine learning tools to detect occurrences of hidden text salting. The authors provide a framework that helps identify known and new hidden text-salting tricks. The authors provide three change detection measures utilising a combination of text features, such as length, edit distance-like measures and complexity, using Kolmogorov complexity utilising Lempel-Ziv compression. A machine learning model is then trained on a dataset of malicious emails labelled as incorporating these hidden text-salting tricks. Emails with a ``new'' trick are presented to the system, and this trick is detected. This provides a robust platform for further research, especially regarding the use of similar techniques on datasets newer than 2008.\par\smallskip

Similar work was later conducted in 2010 by Moens et al. \cite{moens_identifying_2010}, expanding on the approach by utilising features of the rendering engine itself to both identify occurrences and reconstruct the ``cover text'' being used by the attackers. The utilisation of the rendering engine to determine both the attacker strategy and the cover is important in the context of further email analysis. Researchers can use this framework to help analyse emails without optical character recognition, a computationally expensive tool in 2010.\par\smallskip

In 2020 \foreignlanguage{spanish}{Jáñez‑Martino} had provided a topic classification technique for email spam\cite{janez-martino_classication_2020}. Work conducted by Murugavel and Santhi (2020) explored the identification of particular threads of malicious email using text analytics \cite{murugavel_detection_2020}. The main output of Murugavel's work was a system to compare these identified spam topics to an existing database of prevalent spam topics to classify a particular email as spam. Wang et al. performed N-gram analyses and topic modelling to analyse spam trends, however, they did not discuss content concealment as being a limiting factor to this analysis\cite{fumera_spam_nodate}.\par\smallskip

Content concealment techniques have been further discussed in the context of attack mechanisms in emails in 2023 \cite{janez-martino_review_2023}. This research applies a system of removing any known content concealment using machine learning classification. The authors make significant contributions towards the classification of spam emails and solutions to content concealment techniques used by hackers. While their system is effective given the dataset used, one of the key concerns of phishing is the evolution of techniques used by attackers, which could result in this trained machine-learning classifier being ineffective against new attacks. \foreignlanguage{spanish}{Jáñez‑Martino} further evaluated existing methods of classifying emails into malicious and non-malicious\cite{janez-martino_review_2023}.\par\smallskip

Work has been conducted on analysing malicious text embedded into images by using computer vision techniques such as optical character recognition (OCR) \cite{fumera_spam_nodate, janez-martino_review_2023, annadatha_image_2018, improved_annadatha_image_nodate, yan_gao_image_2008} and further analysis into the applicability of adversarial machine learning techniques to deceive these OCR-based filters \cite{song_fooling_2018, imam_ocr_2022}. A significant portion of this work shows promising results for the use of OCR for classifying image-only email datasets as malicious or not, yet the use of these systems to facilitate the detection of content concealment was not discussed.\par\smallskip

van Dooremaal et al. analyse both malicious and benign websites using combined feature extraction of both visual and text-based representations of content\cite{van_dooremaal_combining_2021}. This research proves that there is still room for expanding content analysis techniques for detecting deceitful content in digital communications, with high applicability to this work on visual representations of emails. Additionally, their meta-analysis of existing papers shows tactics such as the use of search engine results for keywords on websites, but such tactics do not generalise to email due to a lack of public search engines for email content and valid senders.

%% file: 4ResearchGap.tex
Previous studies surrounding malicious emails have predominantly focused on improving efforts to classify existing attacks or identifying patterns for detecting new attacks. While these contributions are crucial for advancing anti-spam and anti-phishing measures in operational systems, a significant research gap remains in addressing content concealment in a way that is resistant to evolving attack methods.\par\smallskip

\noindent We have identified the following research gaps:
\begin{itemize}[left=5px]
    \item \textbf{Recipient View}: Previous research applies rendering and optical character recognition (OCR) processes primarily to identify attacks hidden in images or to examine how attackers restructure emails. However, there is a lack of comprehensive approaches that focus on understanding what the email recipient actually sees, especially in the context of content concealed from the recipient, but visible to email filters.
    \item \textbf{Attacker Strategies}: Research focusing on hidden text salting or content concealment techniques generally aims to identify the occurrence of these attacks rather than providing a detailed understanding of the attackers’ goals and strategies. This gap hampers the development of more effective countermeasures.
\end{itemize}
\noindent To address these gaps, this study aims to answer the following research objectives:
\begin{itemize}[left=5px]
    \item \textbf{Identify Content Concealment}: By analysing emails from the perspective of both the recipient and the mail filter, we can identify potential occurrences of content concealment for further analysis.
    \item \textbf{Explore Content Concealment sub-typess}: Next, we explore emails with a high likelihood of concealment spread across a variety of strata, with the purpose of uncovering a broad range of sub-types attackers use when concealing email content to bypass mail filters. 
    \item \textbf{Mechanistic Understanding of Content Concealment}: Furthermore, by extracting the specific CSS tricks used by attackers to conceal content within emails, we can guide further research into the key areas to be targeted by countermeasures to these attacks.
\end{itemize}

%% file: 5Methodology.tex
This section covers the selection and justification of datasets, the preprocessing steps applied to the data, the hierarchical stratified sampling method used to select a representative sample, and the manual analysis conducted to identify instances of content concealment in emails.
The primary challenge overcome throughout this methodology is the sparse and inconsistent nature of emails, including challenges with rendering and analysis.\par\smallskip

\subsection{Dataset Selection and Justification}
The dataset selection process is critical for analysing the prevalence of concealed content in emails. 
As not every malicious email uses content concealment, a source of malicious emails containing a large volume of unique emails would be beneficial to identifying patterns in sub-types and CSS tricks.
Additionally, attacker methodologies change over time, making it beneficial to have emails representing the attack landscape over an extended period.
Because we are primarily concerned with sub-types that attackers use to avoid mail filters, each email example must be exactly what was intended by the sender.\par\smallskip

The Spam Archive dataset \cite{guenter_untroubledorg_nodate} collected by Bruce Guenter contained 8,419,559 spam emails from March 1998 until April 2024. Each email is stored and redistributed precisely as the destination mail server received it, with no modification by inbound mail servers or through forwarding by reporting parties. 
This dataset has collects malicious emails using a series of honeypot email addresses, and has a large frequency of collected emails between 2003 and 2018. In 2018, one of the domain names used by the honeypots was allowed to expire, which resulted in a significant decrease in email quantity after this point. 
Despite this, having a consistent source of emails stored and disseminated in the RFC5322-compliant Internet Message Format enables researchers to see exactly how the email was received. 
Due to the low overall quantities of emails before 2003 and after 2018, we will be specifically targeting this period, with 8,255,071 total emails.
We studied this dataset principally because, by design, it contains unsolicited emails in their as-received form. This is in contrast to other public email datasets, which may contain self-reported examples of spam and often include modifications made through forwarding or inbound mail filters.
\par\smallskip

\subsection{Data Preprocessing}
The general requirements for detecting concealment in an email are that the email may have content that cannot be seen by recipients and that our analysis techniques can render the email how the sender intended it to be viewed by the recipient.
We undertook significant filtering and processing steps to reduce the dataset only to emails that will allow us to understand content concealment in emails.\par\smallskip

\begin{enumerate}[label=\textbf{(\roman*)}, left=-10px]
\item \textbf{HTML rendering}\\
For each email in the dataset, we determine if there is a possibility for content concealment in that email. 
Emails may either have a single content type or use MIME to contain a multipart message with more than one content type. 
One of the more prevalent content types is ``text/plain''; the mail client renders the text of the email with no sender-specified markup occurring. 
Emails which have this content type as their only part are not subject to content concealment and are not the subject of this study.
Instead, only email parts with \texttt{text/html} as their content type have the ability to conceal content when rendered to the recipient.
For MIME-multipart emails, we first determine if \texttt{text/html} is a content type of one of the parts of this email and next that the \texttt{text/html} part is not an attachment to the email.
If this is the case, the email has the potential for content concealment to occur.
In the dataset, 4,955,370 emails do not contain HTML as a rendered part and are, therefore, unable to conceal content. This leaves us with 3,299,701 emails we can further assess.\par\smallskip

\item \textbf{No remote content}\\
Emails using HTML allow images to be inserted in multiple ways. These include attachments, Content-ID referenced parts, remote content and base-64 encoded images. 
While each has pros and cons, typically involving compatibility with the mail client, most of these methods involve sending the entire image within the email to the recipient.
Remote content is the outlier here, where the sender can host an image on a web server, and this image is retrieved by the recipient's email client when the email is opened.
As discussed earlier, the only place an email can be considered to be what was intended and possible by the sender of an email is when the destination mail server receives it.
Emails which rely on remotely loaded images are subject to change over time and place.
The sender may enforce Content Distribution Network (CDN)-like distribution of the image, such as making it only available to IP addresses owned by the ISP of the intended recipient or changing which image is loaded by concerned parties, such as ISPs, mail service providers or other groups.
Additionally, malicious images are subject to being removed by the hosting organisation after a short period of time, resulting in future attempts to access the remote content being inaccurate as to what the sender initially tried to show to the recipient.
Considering the difficulty in acquiring these remote images, studying emails not in real-time presents a difficulty to researchers because the images cannot be expected to be authentic.
As such, we do not study emails which rely on remotely loaded images, as the rendered version of the email could be inconsistent with the initial attempt by the attacker.
1,260,292 emails used remote content and thus are unable to be assessed. 2,039,409 emails are static and can be assessed without concern of remote content having been modified.\par\smallskip

\item \textbf{English Only}\\
While spam and phishing are concepts that transcend language and culture, we limit our study to email in English. This is because we only have the skill to conduct the manual analysis step in English. Also, limiting the study to one language also allows us to use a single Optical Character Recognition (OCR) process.  We use the fastText lid.176 model\cite{joulin2016fasttext, joulin2016bag}, a fast language identification model.
While this does pose a limitation to the work, most of the specific content concealment strategies being explored are part of HTML and CSS, which generalise well across languages.
662,705 of the emails left were identified by the language identification model as not being in English.
Russian (262,756 occurrences), Japanese (201,662 occurrences), Chinese (49,910 occurrences) and Spanish (24,069 occurrences) make up the majority of the other languages identified.
This results in 1,376,704 emails in English and, therefore, suitable for further processing.
\par\smallskip

\item \textbf{No encoding errors}\\
Next, we consider fringe cases about the encoding of the emails in the dataset.
Some emails would cause errors during loading or rendering later in the process.
While significant steps were taken to analyse this issue, a certain portion of emails cause difficulty by purporting to be in one transfer encoding but contain bytes that are not renderable in that specified transfer encoding.
Further encoding errors about character encodings and fonts were detected at later stages of email processing, and these emails were also discarded.
As a result, all emails identified with bytes that are not renderable by the transfer encoding specified in the email or are manually identified as not being rendered correctly are filtered out.
Hence, we do not consider emails with encoding errors to be within the scope of this work.
34,732 emails were identified as having encoding errors, resulting in 1,341,972 emails which had no encoding errors and could be further processed.
\par\smallskip

\item \textbf{No IE/MSO directives}\\
Another key feature of HTML in emails is the use of comments, which are portions of the HTML that are to be ignored by the renderer in most circumstances.
Software such as Microsoft's Internet Explorer or Microsoft Outlook uses HTML comments in a non-standard way, which results in some parts of the email body being selectively rendered, even if contained in a comment.
These directives allowed the sender to provide content in an HTML comment specifically rendered only by the intended client, such as Outlook.
This allows users sending content between Microsoft clients to have features which are not standard across other clients, such as Gmail.
As these directives may cause emails to have multiple versions, they increase the complexity of identifying content concealment.
As such, we do not further analyse any emails which contain these directives to ensure there is no ambiguity in the rendering of each email.
134,686 emails contained IE/MSO directives, leaving 1,207,286 emails that had no client-specific rendering features and are thus suitable for further processing.\par\smallskip

\item \textbf{CSS availability}\\
The HTML renderer we use is Chromium, an open-source web renderer which is used in many web browsers and mail clients.
This allows the rendered version of the email to match the way the email may have originally been rendered to a potential recipient.
Despite this, many mail clients have further restrictions over the web-based Chromium engine, such as disabling Javascript.
One example of CSS often available in Chromium-based web clients but not mail clients is CSS animations, which work to animate content on the screen without directly utilising Javascript code in the source of the content.
Many mail clients do not allow CSS animations, yet the Chromium web engine would allow this CSS animation to be performed, hindering our OCR process.
Work was taken to align the rendering properties of the Chromium agent with the features available to most mail clients.
However, differences between all mail client features prevent this from being completely accurate.
This was not a direct filtering step, and emails were retained even with invalid CSS, but that CSS was modified so it would not be utilised by the renderer.\par\smallskip
\end{enumerate}
\input{latex-figures/sankey_filtering}
After preprocessing the data, we are left with 1,207,286 emails which have the ability for concealment to occur and the potential to be accurately detected by our study.
Figure \ref{fig:sankey_filtering} displays the total number of emails in the dataset and the quantities removed during each stage of the preprocessing steps.

\subsection{Rendering and Tokenisation}
One of our primary research objectives is to identify and analyze content concealment sub-types in emails. To achieve this, we required a method of comparing what the user sees in an email to the raw source of the email.
For this purpose, we developed a dual-perspective approach to inspecting each email.\par\smallskip

The two perspectives we would like to compare are the Mail Filter View, how a theoretical email filter may interpret the text in an email, and the Recipient View, the words that would have been rendered to the recipient of the email using a standard email client.
Mail Filter View involves extracting text from an email source by stripping all HTML and CSS and tokenising the resulting text.
Recipient View utilises publicly available HTML rendering tools to render a version of the email similar to what was originally intended in the email source.\par\smallskip

Here we detail the exact processing steps involved for each perspective.

\subsubsection{Mail Filter View}
In normal circumstances, HTML and CSS would be used to structure the body of an email and provide a mechanism for the sender to specify rendering aspects to the recipient, such as font size, font colour, background colour and text position.
Assuming some mail filters intend to analyse the text that is rendered to the recipient, we need a way to extract just text which is eligible to be rendered from the email source.
Commercial mail filters typically do not provide the exact details of the email processing steps, so instead we try to find an effective way to extract the text contained in the raw source of the email while removing any non-renderable content.
This eligible text would be anything from the \texttt{text/html} part of the email, which is not an HTML tag, comment, or CSS.
We use the Python library BeautifulSoup\cite{beautifulsoup} to extract this eligible text.
Next, HTML tags, punctuation and non-renderable characters, such as line breaks, are removed and used as separators between words in the text.
We observe that due to the ability of CSS to reposition elements around the render area, we cannot easily infer the relative position of text on a renderer, especially with the impact of differing display settings on end-user devices. 
Because of this, the result from this Mail Filter View is a list of words in the logical order they appeared in the source of an HTML part of an email.

\subsubsection{Recipient View}
For the second pathway, we focus on extracting only the text which should have been shown to the recipient of an email.
The email content is then rendered using Playwright\cite{playwright}, a Python library used to automate browser interactions and typically used for automated testing of web-based services.
Although not all email clients have the same functionality, the constant change in feature availability across email clients prevents complete rendering accuracy, and instead, we must go for an approach using a client with the most supported features.
We render using a headless Chromium browser, which supports most features expected from the typical email renderer.
Once the email has completed rendering, we take a full-page screenshot to represent what the user would have seen in the email.
Then optical character recognition (OCR) is performed using PPOCR\cite{du_pp-ocrv2_2021} to extract the text from the full-page screenshot.
PPOCR extracts bounding boxes for blocks of text, providing both the text extracted and the spatial positioning of this text.
Further, while the spatial positioning of text could be important in future because it is something we are not yet extracting from the Mail Filter View, we ignored this output.
The text extracted by PPOCR was stripped of stop characters, punctuation, and any non-renderable characters.
The end result from Recipient View is a list of words that should have been displayed to the recipient of the spam email.\par\smallskip

\subsubsection{Feature Analysis}
We now have two lists of words, one which closely matches what a mail filter might extract when analysing the body of an email and another which closely matches what would be seen by the recipient of the email.
In occurrences of content concealment, we should expect to see some differences between these two lists of words.\par\smallskip

In the existing literature, some different metrics have been used for comparing the rendered version of an email to its source, such as Tolerant Overlap Coefficient (TOC) and Kolmogorov Complexity \cite{bergholz_detecting_nodate}.
TOC measures the number of tokens that have an edit distance below a threshold between two sets, which is beneficial in determining when entire words are added or removed from the text and handling errors in OCR or tokenisation to a small extent.
While TOC is a beneficial metric for determining the difference between two lists of words including potentially for the detection of content concealment, there is one notable drawback. 
TOC will ignore the concealment for cases with insertions or deletions in a word that do not exceed the tolerance threshold. A well-tuned implementation of TOC may perform well at this task overall. However, that is not the immediate subject of this study.\par\smallskip

Kolmogorov complexity functionally examines the shortest possible description for some input given a fixed universal description language, or in other words, the compression ratio of the input.
We could compare the Kolmogorov complexity\cite{kolmogorov1968three} of two strings using a practical compression algorithm like Lempel-Ziv\cite{ziv1977universal} to determine how much more information is required to represent one string over another, representing the total amount of new information in that string.
One concern with applying Lempel-Ziv is that it depends on the ordering of the input, which we have already determined to not be reliable relative to email source and its rendered text.\par\smallskip

Instead, we are focusing on Jaccard distance, the inverse of the Jaccard Similarity Coefficient or Jaccard Index\cite{jaccard1901etude}. This set function measures the similarity between finite sample sets, defined as one minus the size of the intersection divided by the size of the union of the sample sets. A Jaccard distance of 0 indicates that there are no differences between the elements in two sets (i.e., the sets are identical). In contrast, a Jaccard distance of 1 indicates that there are no matching items between the two sets (i.e., the sets are completely dissimilar).\par\smallskip

We transform each of our email views into sets, with each element being a word in the list extracted earlier in the methodology. In doing so, Jaccard distance can identify specific insertion and deletion of certain words between the two versions of the email and compare when tokens are inconsistent between each email view, presumably due to content concealment. \par\smallskip

We conducted some preliminary analyses during which we explored a small test group of emails containing concealment, during which we identified Jaccard distance as a suitable metric to suggest when concealment may have occurred.
Because we are trying to sample a large dataset to find examples of content concealment in emails, Jaccard distance might allow for a targeted selection of emails with a high possibility of concealment. This assumption was confirmed in our results, section \ref{sampling_effectiveness}.

\subsection{Stratified Sampling}
To effectively identify content concealment sub-types in emails, our methodology requires selecting emails that are capable, as described above, of containing concealed content. 
Despite evolving countermeasures, this selection must represent a diverse range of attacker methodologies. 
To achieve this, we utilized hierarchical stratified sampling\cite{cochran1977}.
We applied stratum-based categorisation across our preprocessed dataset, allowing us to sample emails from each sub-stratum, resulting in a sample that covers a wide range of values in each metric rather than being biased by density.
By ensuring representation across each stratum, we can better analyze variations and patterns within different segments of the dataset that we may not see through simple random sampling.\par\medskip

Specifically, we used three primary strata: time, Jaccard distance and HTML length. 
These strata were further divided into levels, allowing for a comprehensive and structured sampling approach.\par\smallskip

\begin{enumerate}[label={}, left=4px, itemindent=0px]
\item \textbf{Time: }
Historical trends in cybersecurity show that attackers continually adapt and refine their strategies to evade detection\cite{noauthor_apwg_2024}.
Stratifying by year allows us to capture the temporal evolution of concealment strategies. This enables a more comprehensive analysis of how these strategies change and provides insights into the historical effectiveness of email filters.
Because the quantity of emails is greatest between the years 2003 and 2018, we choose these 16 years as the 16 levels within the time stratum.\par\smallskip

\item \textbf{Jaccard Distance: }
We hypothesise that concealment may occur with both large and small differences between the two email views.
The purpose of utilising a difference metric such as Jaccard is that it directly represents the ratio of content which does not appear in both views of the email.
By stratifying based on Jaccard distance, we ensure that sample of emails covers any potential range of quantity of concealed content.
This is crucial for analyzing the effectiveness of concealment strategies and the performance of filtering algorithms across different levels of concealment. 
We include 5 levels for the Jaccard distance, ensuring representation from emails with minor differences between views to emails with almost no similarity between views.\par\smallskip

\item \textbf{HTML Length: }
We further hypothesised that the strategies utilised for content concealment may vary between long and short emails. 
Additionally, the Jaccard distance calculation is influenced by HTML length, with shorter emails potentially having a greater Jaccard distance for the same overall quantity of concealed content compared to longer emails. 
By including HTML length as a stratum, we ensure that the sample captures potential differences in concealment sub-types and CSS tricks used in long versus short emails.
This helps isolate the effect of content concealment on the Jaccard distance from the effect of HTML length and allows for a more nuanced analysis of concealment across different HTML lengths. \par\smallskip

We include 5 levels for HTML length, evenly distributed between the shortest email with a valid \texttt{text/html} part and the 95th percentile in length. 
Emails with HTML parts longer than the upper 95th percentile are outliers with abnormally long bodies which posed significant rendering difficulties.
\end{enumerate}\par\medskip

\noindent We label each email based on the aforementioned strata, a total of 160 sub-strata. 
Our target was to have a total sample of 1000 emails, which would require approximately 7 emails sampled per sub-strata.
Because some sub-strata did not contain 7 total emails, we had to re-sample our sub-strata for additional unique emails.
Preliminary analysis of the sampled emails indicated that errors with the encoding of the content body resulted in some emails having incorrect language detected by fastText in the preprocessing stage.
We also identified some issues with the rendering and OCR process that resulted in the emails being unsuitable for manual analysis and having to be discarded.
In the end, we aimed to sample up to 13 emails per sub-strata, with a total of 1483 emails.
Of these 1483 emails, 499 emails suffered from either language detection or processing errors, resulting in a final sample of 984 emails.
\input{latex-figures/histogram_jaccard}

\subsection{Manual Analysis of Sampled Emails}
We created a custom email browser to analyse four separate perspectives of each email.
We used the email's raw source, a rendered version of the HTML in the email, the Mail Filter View and the Recipient View.
Our tool utilised Python's difflib library\cite{python-difflib}, specifically the ndiff tool, to visually highlight differences between the Mail Filter View and Recipient View.
This difference allowed for instant inspection of the rendered email to determine whether content was indeed hidden, and further analysed the raw source of the email to determine the concealment sub-types and CSS tricks used.

\begin{enumerate}[label={}, left=4px, itemindent=0px]
\item \textbf{Concealment Criteria: }
During our first analysis of the sample emails, we identified whether or not each email had any content concealed from the user.
This binary classification involved visually inspecting the difference between the Mail Filter View and Recipient View to determine if any content was different, and for each difference, the raw source of the email was inspected to identify whether or not it was an issue with the OCR stage or indeed content concealment.
To make this judgement, we determined that if text was identified in the source of the email but not in the rendered view but would be rendered without the presence of HTML or CSS, it was an indication of concealment having occurred.

\item \textbf{Judgement and Classification: }
For the second analysis of the sampled emails, we focused only on emails which had concealment identified in the first analysis.
We had two key criteria to inspect for.
The first is to determine what part of the email was being modified, for example, whether it was a single word or a large paragraph being concealed.
The second is to utilise the raw source of the email to identify which CSS tricks contribute to the content concealment, for example, font colour or font size.
We further utilised the rendered version of the email to examine exactly what the recipient would have seen and determine what impact the utilised CSS tricks have on the rendering.
\end{enumerate}

%% file: latex-figures/sankey_filtering.tex
\begin{figure*}[t]
\centering
Email Filtering Quantities\par\medskip
\includegraphics[width=1\textwidth]{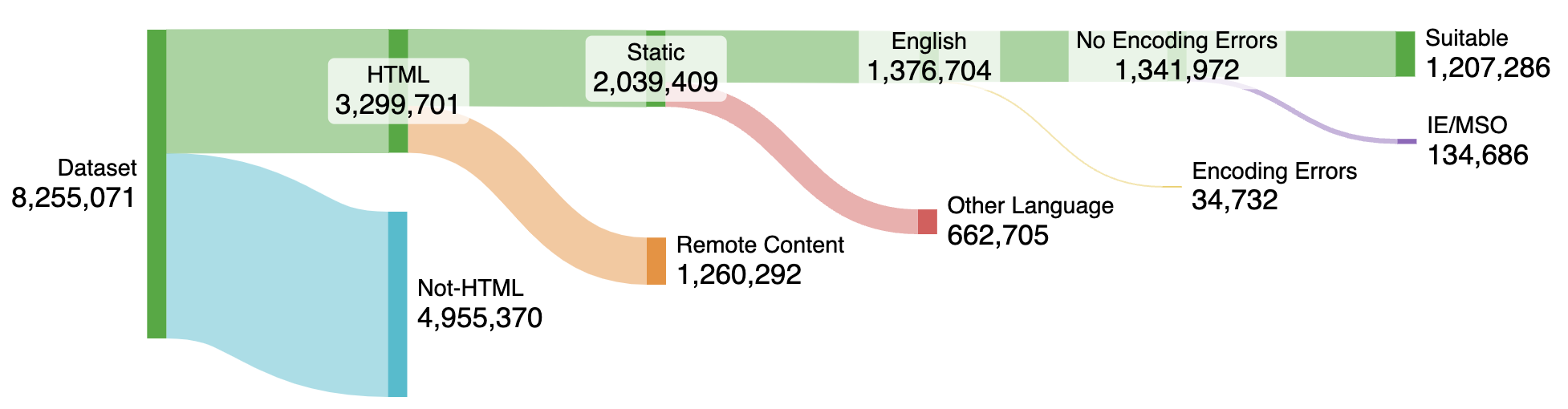}
\caption{Sankey diagram highlighting the quantities of emails filtered out during each preprocessing stage of the methodology.}
\label{fig:sankey_filtering}
\end{figure*}

%% file: latex-figures/histogram_jaccard.tex
\begin{figure}[bt]
\centering
\begin{tikzpicture}
\begin{axis}[
    ybar,
    bar width=12pt, 
    ylabel={\# Emails},
    xlabel={Jaccard Index},
    ymin=0,
    enlarge x limits={abs=0.75cm}, 
    xtick={1, 2, 3, 4, 5},
    xticklabels={0.0-0.2, 0.2-0.4, 0.4-0.6, 0.6-0.8, 0.8-1.0}, 
    xticklabel style={rotate=90, anchor=near xticklabel},
    nodes near coords,
    nodes near coords align={vertical},
    every node near coord/.append style={font=\scriptsize, color=black, anchor=south}, 
    title={Concealment Occurrences by Jaccard Index},
    ]
    \addplot +[
        ybar, 
        bar shift=-6pt, 
        style={fill=red!20, draw=red, mark=none, postaction={pattern=north east lines, pattern color=white}},
        nodes near coords,
        visualization depends on=y \as \rawy,
        nodes near coords align={vertical},
        every node near coord/.append style={font=\scriptsize, color=black, anchor=south},
    ] table [x=jaccard_bin, y=with_concealment, col sep=comma, meta index=1] {latex-figures/concealment_by_jaccard.csv};
    
    \addplot +[
        ybar, 
        bar shift=6pt, 
        style={fill=blue!20, draw=blue, mark=none, postaction={pattern=dots, pattern color=white}},
        nodes near coords,
        visualization depends on=y \as \rawy,
        nodes near coords align={vertical},
        every node near coord/.append style={font=\scriptsize, color=black, anchor=south},
    ] table [x=jaccard_bin, y=no_concealment, col sep=comma, meta index=2] {latex-figures/concealment_by_jaccard.csv};

    \legend{Concealment, No Concealment}
\end{axis}
\end{tikzpicture}
\caption{Grouped bar chart showing the number of emails where content concealment was detected by Jaccard index in our sample.}
\label{fig:histogram_jaccard}
\end{figure}
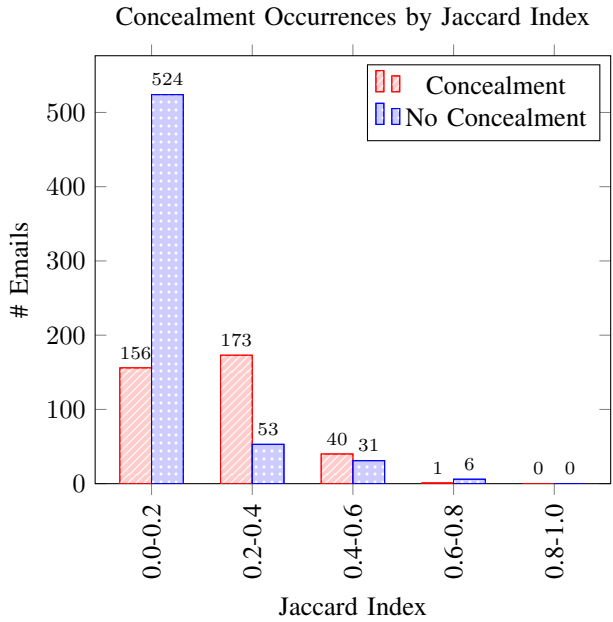

%% file: 6Results.tex
\input{latex-figures/histogram_year}
This results section presents the findings of our analysis of email content concealment. We first report the detection of concealment sub-types used by attackers. Next, we explore the CSS tricks that enable these sub-types and the relationship between the sub-types and tricks.
\subsection{Sampling Effectiveness}  \label{sampling_effectiveness}
As mentioned, our sample from the chosen dataset contained 984 emails. 
After manual analysis, we discovered that 370 of these emails contained concealed content, while 614 did not.

In Figure \ref{fig:histogram_year}, we illustrate the distribution of emails with and without concealment across our time stratum.
The change in attacker strategies over time displayed in Figures \ref{fig:histogram_yearly_tactics} and \ref{fig:histogram_yearly_techniques} also display how the attacks change over time, further reinforcing the importance of sampling over a time stratum.
\input{latex-figures/boxplot_length}

The majority of emails in the dataset had a Jaccard distance between 0 and 0.4. We noticed that our sample has underrepresented emails with high Jaccard distances, with most of our sampled emails in the 0-0.4 range. 
Because Jaccard distance is essentially a ratio covering the amount of similarity between our two views, and each email source must contain the content being displayed to the recipient, it is unlikely to have Jaccard distance values that are large.
For example, a Jaccard distance of 0.5 indicates that the size of the intersection of the words in each email view is half the size of the union of those two views.
Because the email source must contain, at minimum, the content which has been displayed to the recipient of an email, we can guarantee that the intersection of the two sets should contain the mail filter view.
Because of this, unless there is significant concealment which is greater than the amount of text rendered to the recipient, we would expect to see Jaccard distances of less than 0.5.
We identified that the majority of emails with either language detection or processing issues during the manual identification stage of the methodology and were discarded from future analysis had these large Jaccard distances, which is why they are not represented in the data in our analysis.
Within our sample, we compared the Jaccard distances of emails with and without concealment. In particular, Figure \ref{fig:histogram_jaccard} highlights the difference in concealment by Jaccard distance level.

We also considered the importance of HTML length as one of our strata. We found that shorter emails tended to have a lower percentage of concealment compared to longer emails, as shown in Figure \ref{fig:boxplot_length}.\par\medskip

\subsection{Concealment Sub-types}
We use the term \textit{sub-type} for the general way content is concealed and the term \textit{CSS Trick} for the HTML/CSS attributes used for the attack, as we describe further below.
During our analysis, we identified three key sub-types attackers use to conceal content in emails. 
Figure \ref{fig:methods_venn_diagram} highlights the overlapping use of each sub-type within the sample.
In the Appendix, we explain an example of each sub-type and how it functions in Figure \ref{fig:concealment_quantities}.\par\medskip

\input{latex-figures/methods_venn_diagram}

\begin{enumerate}[label=(\roman*)]
\item \textbf{Add Paragraph}\\
For this sub-type, attackers introduce large portions of text and other HTML into the body of the email, which the recipient cannot see.
Examples of this may include passages from books, scraped webpages and more. 
Additionally, this sub-type varied in size, where some attacks had significant quantities of concealed content, whereas others had as few as 20 characters concealed in bulk. 
As this sub-type allows for all concealed words to be dictionary words, the ability for simple mail filters to easily ignore them may be limited.
This was the most prevalent sub-type identified, with 214 emails that had concealed added paragraphs. 43 of these emails had utilised at least one other attack sub-type.
A direct example is provided in the Appendix, Figure \ref{fig:example_add_paragraph}.\par\medskip

\item \textbf{Disrupt Word}\\
For this sub-type, attackers specifically target spam-like words in their email, which may otherwise result in detection by mail filters.
Typical examples of this attack involve adding single or multiple characters throughout spam-like words in the email body without these additions being visible to the recipient. 
In doing so, the attacker can change the overall meaning of a sentence from the perspective of a mail filter which interprets known sentences. 
One example of how this sub-type would affect mail filters is that high-risk words could be presented to the recipient of an email, even if they are obfuscated or otherwise modified when seen by the mail filter.
We discovered 186 emails that employed disrupted words, with 47 utilising at least one other attack sub-type.
An example is provided in the Appendix, Figure \ref{fig:example_disrupt_word}.\par\medskip
\input{latex-figures/histogram_yearly_tactics}

\item \textbf{Insert Word}\\
We discovered that attackers would insert entire words into the email to disrupt sentence flow.
In this attack, sentences with high-risk phrases had words added to them that were concealed from the recipient.
One impact of this strategy is that it could rely on as few as 2 additional HTML tags per sentence and not have any non-words.
We discovered that this attack is not only the least prevalent as per Figure \ref{fig:methods_venn_diagram}, but it was also only found in emails in the earlier years of the sample, with no occurrences after 2011, as seen in Figure \ref{fig:histogram_yearly_tactics}. 
Our first hypothesis for this finding is that this sub-type was discovered by attackers not to be as effective as the other two sub-types identified.
The second is that the attack relies heavily on the context of the sentence and the word being inserted, as inserting words into a sentence to prevent detection could require intimate knowledge of the countermeasures themselves. 
Because of this, there is potentially a need for burdensome manual intervention to perform this concealment sub-type successfully.
We discovered 23 emails containing inserted words, with 15 of these utilising at least one other attack sub-type.
An example is provided in the Appendix, Figure \ref{fig:example_insert_word}.\par\medskip
\end{enumerate}
We note that both the "Disrupt Word" and "Insert Word" sub-types may yield Mail Filter View and Recipient View word lists that only differ by small amounts, which would account for concealment existing despite showing low Jaccard scores.

\subsection{CSS Tricks}
We further explored the CSS tricks used to conceal content in these emails by manually identifying the CSS tags applied to the concealed content.
We group the specific CSS tricks used to conceal content into 5 main categories. In Figure \ref{fig:upset_plot_techniques}, we demonstrate the quantities of each CSS trick identified as contributing to concealment in an email. Figure \ref{fig:histogram_yearly_techniques} further illustrates the relative use of each CSS trick per year, noting that the CSS tricks appear to vary in use but are present throughout the entire dataset. One noteworthy feature in Figure \ref{fig:histogram_yearly_techniques} is that 2010 appears to have more use of CSS tricks, but Figure \ref{fig:histogram_year} shows that there are only a few more total occurrences than in the next most significant year. This is because the emails in the sample from 2010 used more combined features, whereas other years may have used more singular features.
Examples of some of the CSS tricks identified are available in the Appendix, Figure \ref{fig:css_techniques}.\par\medskip

\input{latex-figures/histogram_yearly_techniques}

\input{latex-figures/upset_plot_techniques}
\begin{enumerate}[label=(\roman*)]
\item \textbf{Font Colour}\\
Font colour was the most prevalent CSS trick that we identified in the sample, with 264 total occurrences. This CSS trick is characterised by text having a font colour with poor contrast against its background colour so that it is not visible on the recipient's screen. This CSS trick seems simple for both attackers and mail filters, however the complexity is compounded by a few factors. First and foremost, the inheritance of CSS attributes from parent HTML elements means that the exact rendering of any piece of text relies heavily on its context. For example, the CSS ``color'' property could be set to ``white'' for a div surrounding an entire email with a white background. Within the div, text that should be visible to the user can override the ``color'' property to ``black'' or another colour with contrast against the background. Text which does not have an overridden ``color'' property will remain white, and not visible to the recipient. The downside of just using font colour for concealment is that the text still occupies space on the screen and may cause other elements to be positioned unusually for the recipient. 

Because of this, attackers frequently use a combination of font colour and other CSS tricks, such as the ones discussed further in this section. As demonstrated in Figure \ref{fig:upset_plot_techniques}, font colour is used alone 90 times and most commonly combined with font size and text position, or just font size, totalling 72 and 57 additional occurrences, respectively.\par\medskip

\item \textbf{Font Size}\\
We identified font size as the second most prevalent CSS trick in the sample, with a total of 218 occurrences. Attackers use this CSS trick to reduce the text size so that it appears less conspicuous on the recipient's screen. Frequently we observed font sizes of 3px or less. In contrast to the font colour CSS trick, font size is not relative to its background or context. It could also be suggested that due to small font sizes being difficult for users to read, it would be uncommon for legitimate communication to use such small font sizes. As such, it may be easier for this CSS trick to be directly identified by mail filters with little in-depth analysis of the context of the text. 

Figure \ref{fig:upset_plot_techniques} shows that font size was not often used alone, with only 40 occurrences. Instead, it was used most commonly in combination with other CSS tricks a total of 178 times. For example, combining font size and font colour CSS tricks allows for text to be made small to reduce its impact on the position of rendered content and reduce its contrast so that it is not visible to the recipient.
\par\medskip

\item \textbf{Text Position}\\
Additionally, we discovered the use of concealment CSS tricks that allow content not to be rendered in its normal position or not be rendered at all, with a total of 97 occurrences. This attacker sub-type includes the common ``float'' manipulation, where the position of the rendered text is set to be different from where it appeared in the HTML source. Also, we have included situations where ``display'' is set to hidden. This is because they both function similarly for concealment by having the text not render in its original place in the HTML. 

Similar to the font colour CSS trick, simply moving the position of text may not conceal it completely from the user. This is made apparent because only 11 occurrences of text position were used for concealment without any other combined sub-types, as seen in Figure \ref{fig:upset_plot_techniques}.

\par\medskip
\item \textbf{Table Manipulation}\\
We also discovered concealment by manipulating tables within our sample, with a total of 63 occurrences. This CSS trick involves using HTML tables to structure the email content in a way that hides certain parts of the text from the recipient. Attackers can manipulate the visibility and positioning of table cells, rows, and columns to achieve this effect. For example, by affecting the size of an entire column or row in a table, the attacker can conceal cells within a table without directly acting on the cells themselves.

Additionally, table manipulation can be used to create overlapping content where the visible text overlays the concealed text, effectively hiding it from view. 

Similar to other CSS tricks, table manipulation is often combined with other CSS CSS tricks to enhance its effectiveness. For example, combining table manipulation with font colour or font size adjustments can ensure that the concealed content remains hidden while maintaining email structure and appearance. As shown in Figure \ref{fig:upset_plot_techniques}, table manipulation was rarely used in isolation, with 57 of the 63 occurrences used in conjunction with another CSS trick.\par\medskip

\item \textbf{Other}\\
On occasion, the concealed content did not fall into one of the main categories and had few other similar examples in the dataset, so it did not qualify as a category of CSS tricks. The concealment identified as other was often found in combination with another CSS trick, but the rendered email did not immediately match what was expected for that CSS trick category. While we identified 33 emails as using other concealment CSS tricks, 14 emails that were identified as containing concealment did not use one of the other four CSS tricks. These 14 consist of unique CSS tricks with only a single example in the sample, or the exact CSS CSS trick was unidentifiable due to improperly structured HTML and CSS
\end{enumerate}

\subsection{Summary of Results}
Our analysis revealed several important insights into email content concealment.
Out of 984 emails analyzed, 370 (37.6\%) contained concealed content, while 614 (62.4\%) did not.
Concealment occurrences varied across years, with underrepresentation in 2003, 2012, and 2018.

The mean Jaccard index for emails with concealment was 0.22, significantly higher than 0.07 for emails without concealment, underscoring its potential utility in detecting concealment. HTML length and year strata ensure that sub-types that have varied over time or depending on the length of the HTML are also represented in the sample.

The most common sub-type was adding concealed paragraphs with 214 occurrences involving concealed large text blocks.
Disrupted words were the second most common, with 186 occurrences, using characters to obfuscate spam-like words.
Inserted words were the least common sub-type, with 23 occurrences, primarily in older emails.
``Insert Word'' declined in use after 2011, while ``Font Colour'' and ``Font Size'' remained prevalent throughout the years.\par\medskip

Font colour is the most prevalent CSS trick, with 264 occurrences of emails lowering the contrast of some text to conceal it against the background.
Font size was used in 218 instances to minimize text visibility.
Text position was involved in 97 cases to alter or hide text placement.
Table manipulation was used to conceal text within table structures.
Unique or unidentifiable CSS tricks were observed in 33 emails.
We also illustrate the concealment sub-type's use of each CSS CSS trick, as seen in Figure \ref{fig:heatmap_tactic_vs_technique}.\par\medskip

These findings provide insight into the strategies employed by attackers to bypass detection from mail filters and provide a foundation for developing more effective email filtering strategies.

\input{latex-figures/heatmap_tactic_vs_technique}

%% file: latex-figures/histogram_year.tex
\begin{figure}[t]
\centering
\begin{tikzpicture}
\begin{axis}[
    ybar stacked,
    ylabel={\# Emails},
    xlabel={Year},
    ymin=0,
    bar width=11px,
    enlarge x limits={abs=0.5cm}, 
    symbolic x coords={2003, 2004, 2005, 2006, 2007, 2008, 2009, 2010, 2011, 2012, 2013, 2014, 2015, 2016, 2017, 2018}, 
    xtick={2003, 2005, 2007, 2009, 2011, 2013, 2015, 2017}, 
    xticklabel style={rotate=90, anchor=near xticklabel},
    nodes near coords,
    nodes near coords align={vertical},
    every node near coord/.append style={font=\scriptsize, color=black, anchor=south}, 
    point meta=explicit symbolic,
    title={Concealment Occurrences by Year},
    ]
    \addplot +[
        ybar, 
        style={fill=red!20, draw=red, mark=none, postaction={pattern=north east lines, pattern color=white}},
        nodes near coords,
        visualization depends on=y \as \rawy,
        nodes near coords align={vertical},
        every node near coord/.append style={font=\scriptsize, color=black, anchor=south, yshift=-0.2cm},
    ] table [x=year, y=with_concealment, col sep=comma, meta index=1] {latex-figures/concealment_by_year.csv};
    
    \addplot +[
        ybar, 
        style={fill=blue!20, draw=blue, mark=none, postaction={pattern=dots, pattern color=white}},
        nodes near coords,
        visualization depends on=y \as \rawy,
        nodes near coords align={vertical},
        every node near coord/.append style={font=\scriptsize, color=black, anchor=south, yshift=-0.2cm},
    ] table [x=year, y=no_concealment, col sep=comma, meta index=2] {latex-figures/concealment_by_year.csv};

    \legend{Concealment, No Concealment}
\end{axis}
\end{tikzpicture}
\caption{Stacked histogram showing the number of emails where content concealment was detected by year in our sample.}
\label{fig:histogram_year}
\end{figure}
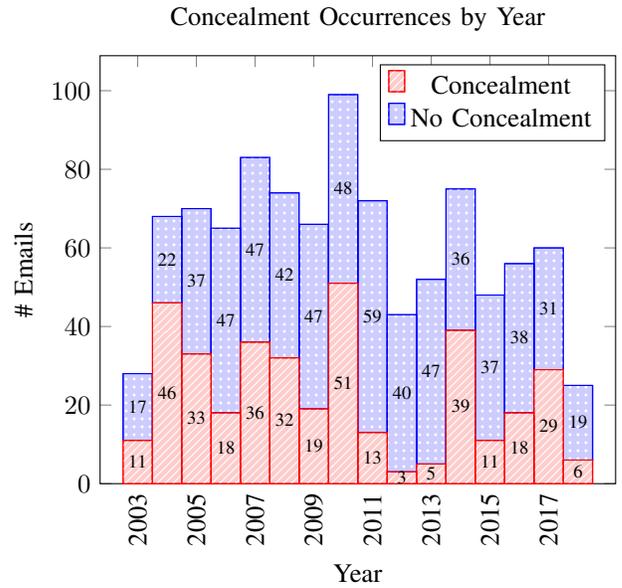

%% file: latex-figures/boxplot_length.tex
\begin{figure}[bt]
\centering
\begin{tikzpicture}
\begin{axis}[
    ybar,
    bar width=12pt, 
    ylabel={\# Emails},
    xlabel={HTML Length},
    ymin=0,
    enlarge x limits={abs=0.75cm}, 
    xtick={1, 2, 3, 4, 5},
    xticklabels={{129-3656}, {3656-7183}, {7183-10710}, {10710-14237}, {14237-17764}}, 
    xticklabel style={rotate=90, anchor=near xticklabel},
    nodes near coords,
    nodes near coords align={vertical},
    every node near coord/.append style={font=\scriptsize, color=black, anchor=south}, 
    title={Concealment Occurrences by HTML Length},
    ]
    \addplot +[
        ybar, 
        bar shift=-6pt, 
        style={fill=red!20, draw=red, mark=none, postaction={pattern=north east lines, pattern color=white}},
        nodes near coords,
        visualization depends on=y \as \rawy,
        nodes near coords align={vertical},
        every node near coord/.append style={font=\scriptsize, color=black, anchor=south},
    ] table [x=part_length_bin, y=with_concealment, col sep=comma, meta index=1] {latex-figures/concealment_by_length.csv};
    
    \addplot +[
        ybar, 
        bar shift=6pt, 
        style={fill=blue!20, draw=blue, mark=none, postaction={pattern=dots, pattern color=white}},
        nodes near coords,
        visualization depends on=y \as \rawy,
        nodes near coords align={vertical},
        every node near coord/.append style={font=\scriptsize, color=black, anchor=south},
    ] table [x=part_length_bin, y=no_concealment, col sep=comma, meta index=2] {latex-figures/concealment_by_length.csv};

    \legend{Concealment, No Concealment}
\end{axis}
\end{tikzpicture}
\caption{Grouped bar chart showing the number of emails where content concealment was detected by HTML Length in our sample.}
\label{fig:boxplot_length}
\end{figure}
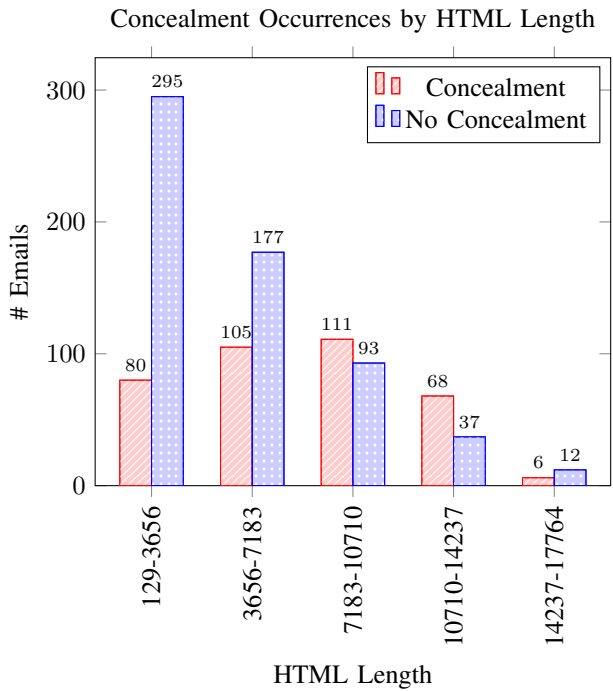

%% file: latex-figures/methods_venn_diagram.tex
\begin{figure}[t]
\centering
Overlap between Concealment Sub-types\par\medskip

\begin{tikzpicture}
    \begin{scope}[fill opacity=0.5]
        \fill[white] (0,0) circle (2cm);
        \fill[white] (2.5,0) circle (2cm);
        \fill[white] (1.25,2) circle (2cm);
        
        \draw (0,0) circle (2cm) node[below left] {};
        \draw (2.5,0) circle (2cm) node[below right] {};
        \draw (1.25,2) circle (2cm) node[above] {};
        \draw (-2.25,-2.25) rectangle (4.75,4.25);
    \end{scope}
    
    \node at (-0.75,0) {Disrupt Word};
    \node at (3.25,0) {Insert Word};
    \node at (1.25,2.75) {Add Paragraph};
    
    \node at (-0.5,-0.6) {139};
    \node at (3,-0.6) {8};
    \node at (1.25,2.25) {171};
    \node at (1.25,-0.5) {9};
    \node at (2.25,1.25) {5};
    \node at (0.25,1.25) {37};
    \node at (1.25,0.75) {1};
    
\end{tikzpicture}
\caption{Venn diagram illustrating the distribution of content concealment sub-types used per email. The diagram shows the overlap between the three main sub-types. The value in each region represents the number of emails that use each concealment sub-type or combination of sub-types.}
\label{fig:methods_venn_diagram}
\end{figure}
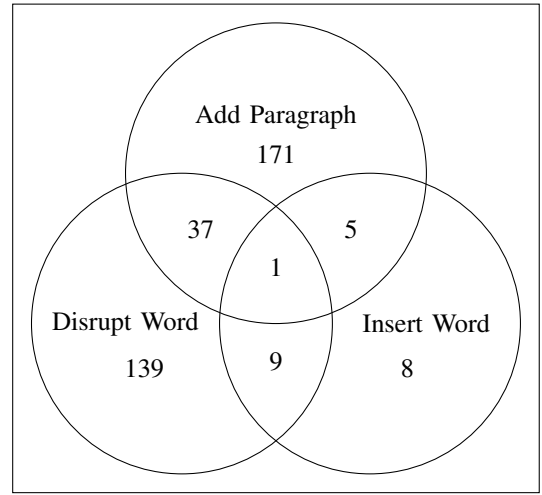

%% file: latex-figures/histogram_yearly_tactics.tex
\begin{figure}[t]
\centering
\begin{tikzpicture}
\begin{axis}[
    ybar stacked,
    ylabel={\# Occurrences},
    xlabel={Year},
    ymin=0,
    bar width=11px,
    enlarge x limits={abs=0.5cm}, 
    symbolic x coords={2003, 2004, 2005, 2006, 2007, 2008, 2009, 2010, 2011, 2012, 2013, 2014, 2015, 2016, 2017, 2018}, 
    xtick={2003, 2005, 2007, 2009, 2011, 2013, 2015, 2017}, 
    xticklabel style={rotate=90, anchor=near xticklabel},
    nodes near coords,
    nodes near coords align={vertical},
    every node near coord/.append style={font=\scriptsize, color=black, anchor=south}, 
    point meta=explicit symbolic,
    title={Concealment Sub-type Occurrences by Year},
    ]
    \addplot +[
        ybar, 
        style={fill=orange!20, draw=orange, mark=none, postaction={pattern=dots, pattern color=white}},
        nodes near coords,
        visualization depends on=y \as \rawy,
        nodes near coords align={vertical},
        every node near coord/.append style={font=\scriptsize, color=black, anchor=south, yshift=-0.2cm},
    ] table [x=year, y=bulk_count, col sep=comma, meta index=2] {latex-figures/concealment_methods_by_year.csv};
    \addplot +[
        ybar, 
        style={fill=yellow!50, draw=yellow!80!black, mark=none, postaction={pattern=north east lines, pattern color=white}},
        nodes near coords,
        visualization depends on=y \as \rawy,
        nodes near coords align={vertical},
        every node near coord/.append style={font=\scriptsize, color=black, anchor=south, yshift=-0.2cm},
    ] table [x=year, y=words_count, col sep=comma, meta index=1] {latex-figures/concealment_methods_by_year.csv};
    \addplot +[
        ybar, 
        style={fill=green!90!lightgray, draw=green!80!black, mark=none, postaction={pattern=crosshatch, pattern color=white}},
        nodes near coords,
        visualization depends on=y \as \rawy,
        nodes near coords align={vertical},
        every node near coord/.append style={font=\scriptsize, color=black, anchor=south, yshift=-0.2cm},
    ] table [x=year, y=insert_count, col sep=comma, meta index=3] {latex-figures/concealment_methods_by_year.csv};

    \legend{Add Paragraph, Disrupt Word, Insert Word}
\end{axis}
\end{tikzpicture}
\caption{Stacked histogram showing the number of occurrences of each sub-type per year. When more than one sub-type was used, we count each occurrence separately.}
\label{fig:histogram_yearly_tactics}
\end{figure}
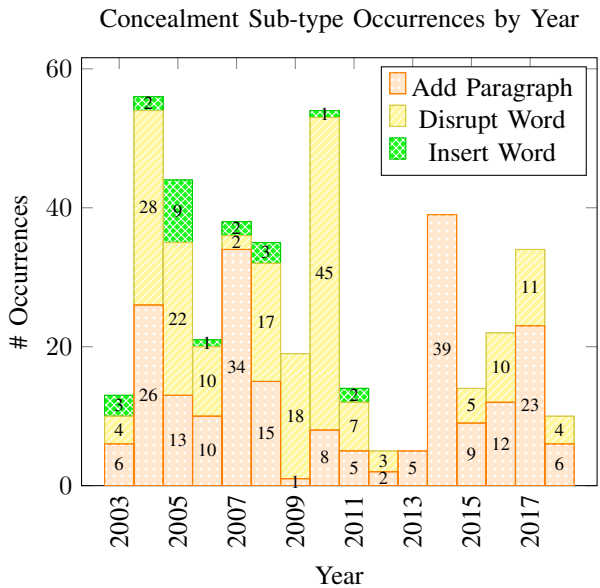

%% file: latex-figures/histogram_yearly_techniques.tex
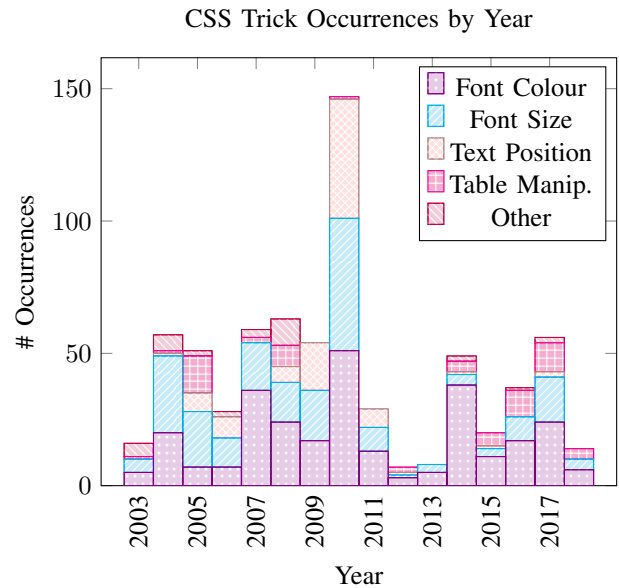
\begin{figure}[t]
\centering
\begin{tikzpicture}
\begin{axis}[
    ybar stacked,
    ylabel={\# Occurrences},
    xlabel={Year},
    ymin=0,
    bar width=11px,
    enlarge x limits={abs=0.5cm}, 
    symbolic x coords={2003, 2004, 2005, 2006, 2007, 2008, 2009, 2010, 2011, 2012, 2013, 2014, 2015, 2016, 2017, 2018}, 
    xtick={2003, 2005, 2007, 2009, 2011, 2013, 2015, 2017}, 
    xticklabel style={rotate=90, anchor=near xticklabel},
    point meta=explicit symbolic,
    title={CSS Trick Occurrences by Year},
    ]
    \addplot +[
        ybar, 
        style={fill=violet!20, draw=violet, mark=none, postaction={pattern=dots, pattern color=white}},
        visualization depends on=y \as \rawy,
    ] table [x=year, y=colour_count, col sep=comma, meta index=2] {latex-figures/css_techniques_by_year.csv};
    \addplot +[
        ybar, 
        style={fill=cyan!20, draw=cyan, mark=none, postaction={pattern=north east lines, pattern color=white}},
        visualization depends on=y \as \rawy,
    ] table [x=year, y=size_count, col sep=comma, meta index=1] {latex-figures/css_techniques_by_year.csv};
    \addplot +[
        ybar, 
        style={fill=pink!50, draw=pink!70!black, mark=none, postaction={pattern=crosshatch, pattern color=white}},
    ] table [x=year, y=position_count, col sep=comma, meta index=3] {latex-figures/css_techniques_by_year.csv};
    \addplot +[
        ybar, 
        style={fill=magenta!40, draw=magenta, mark=none, postaction={pattern=grid, pattern color=white}},  
    ] table [x=year, y=table_count, col sep=comma, meta index=4] {latex-figures/css_techniques_by_year.csv};
    \addplot +[
        ybar, 
        style={fill=purple!20, draw=purple, mark=none, postaction={pattern=north west lines, pattern color=white}},
    ] table [x=year, y=other_count, col sep=comma, meta index=5] {latex-figures/css_techniques_by_year.csv};
    \legend{Font Colour, Font Size, Text Position, Table Manip., Other}
\end{axis}
\end{tikzpicture}
\caption{Stacked histogram showing the number of times a CSS trick was used in an email with concealment per year. When more than one CSS trick was used in an email, we count each occurrence separately.}
\label{fig:histogram_yearly_techniques}
\end{figure}

%% file: latex-figures/upset_plot_techniques.tex
\begin{figure*}[t]
\centering
Concealment CSS Trick Quantities\par\medskip
\includegraphics[width=0.9\textwidth]{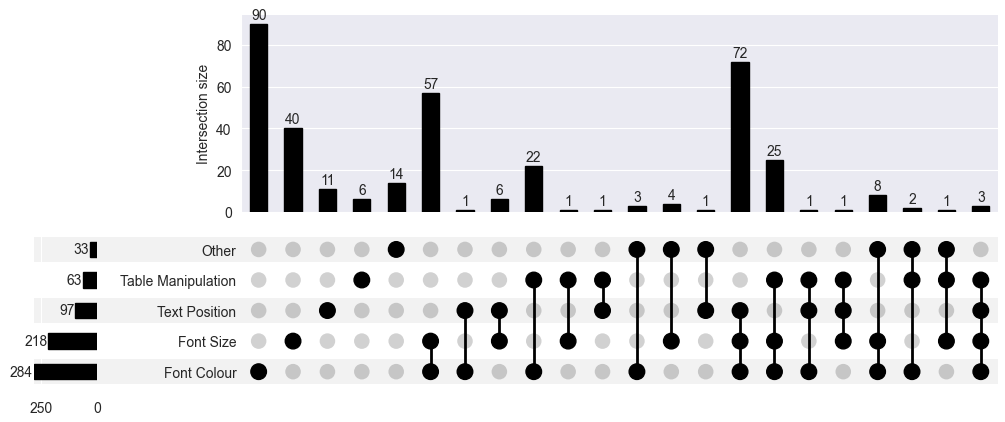}
\caption{This figure illustrates the number of emails where either a single or combination of multiple CSS tricks were used to conceal content. In the figure, rows illustrate the total number of times a CSS trick was used, and the vertical bar graph illustrates when a combination of CSS tricks was used, where the matrix illustrates which CSS tricks are used in that combination. We can see that font colour was the most used CSS trick, combined with other CSS tricks such as font size.}
\label{fig:upset_plot_techniques}
\end{figure*}

%% file: latex-figures/heatmap_tactic_vs_technique.tex
\begin{figure}[t]
\centering
Heatmap of CSS Tricks by Concealment Sub-types\par\medskip
\includegraphics[width=0.5\textwidth]{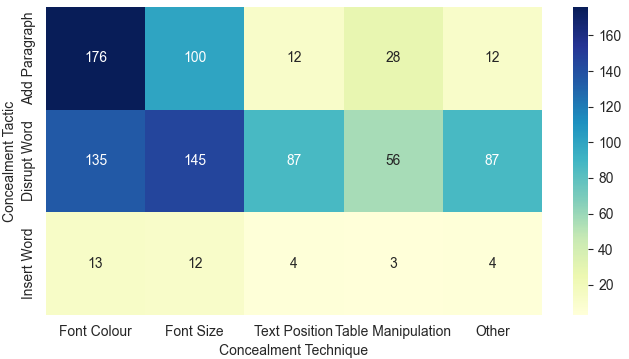}
\caption{This heatmap represents the frequency of occurrences of each CSS trick per concealment sub-type in the sample.}
\label{fig:heatmap_tactic_vs_technique}
\end{figure}

%% file: 7Discussion.tex
CSS provides significant flexibility to the presentation of emails, which is heavily relied upon for tasks such as digital advertising. One of the fundamental aspects of CSS is inheritance. 
Styles applied to HTML elements can be affected by multiple layers of CSS, each applied from different sources. This makes it challenging to predict an element's visual rendering just by statically analysing the HTML source.\par\smallskip

This inheritance is what most HTML rendering engines perform when rendering web pages or emails. 
However, these rendering engines dynamically interpret the content, determining the appropriate visual representation of an element by applying all CSS properties that relate to it. 
The rendering engines also do not inherently care how the content is rendered. For example, an element containing red-coloured text may inherit no background colour. 
Positioning this text on top of a different red-coloured element would result in a human not seeing that text despite it being fully rendered by the renderer.\par\smallskip

Creating a tool that can inherently understand the complexities of CSS rendering and the contextual analysis of what the user would or would not see is a complex task. 
Furthermore, training a machine learning model to detect occurrences of previously unseen concealment sub-types accurately would require that filter to perform this contextual analysis.
This is the core reason behind our selection of OCR, as it allows viewing the rendered email as it would be seen by the recipient.\par\smallskip

Due to the computational cost, the OCR process may not be suitable for real-time analysis of emails in practical mail filtering systems.
Instead, its use as a research tool to understand the concealment sub-types and CSS tricks employed by attackers may shed light on methods to prevent content concealment without the use of OCR.\par\smallskip

In this study, we directly sampled emails between 2003 and 2018 inclusive. In the years since then, there has been a rise in the prevalence and functionality of generative AI tools such as large language models (LLMs).
These tools greatly reduce the barrier to entry for attackers to craft highly convincing malicious emails when presented to users, and we rely on mail filters to prevent these emails from reaching their destination.
We foresee that without addressing the ability of attackers to conceal content in emails, the adversarial capabilities of these generative AI tools could allow attackers to craft convincing emails to the recipient and bypass the mail filter through concealment. The combination of these would allow attackers to repeatedly send the same rendered email multiple times, using machine learning tools to restructure its appearance to the mail filter.
This could allow for attacks that automatically adapt to mail filters, with little intervention required by users.

%% file: 8Conclusion.tex
This study investigated the content concealment sub-types used by attackers in email communication. 
Our objectives were to describe the scale of content concealment and to identify the specific concealment sub-types and CSS techniques employed for concealment in order to understand their impact on email filtering mechanisms. 
This work goes beyond previous studies by analysing a large dataset to determine content concealment, utilising OCR for detecting concealment within text rather than just within images, and detailing the sub-types and techniques used by attackers to deceive email filters.\par\smallskip

We introduced a dual-perspective approach to analysing emails, with the express purpose of identifying potential occurrences of content concealment for further analysis.
We used this approach to sample a large-scale dataset of malicious emails and manually analysed emails for content concealment.\par\smallskip

Our analysis revealed several important insights. 
We found that 37.6\% of the emails in our sample contained concealed content. 
In these emails containing concealment, we identified three primary sub-types used by attackers, including adding paragraphs, disrupting words and inserting words.
We further highlighted the use of various CSS techniques to conceal content, including font colour, font size, and text position, among others. 
These techniques were often used in combination to enhance their effectiveness.\par\smallskip

As has been previously mentioned, one significant limitation of this work is the ambiguous or incorrect encoding present in a large portion of emails in the dataset. 
It was possible for specific concealment sub-types to remain unnoticed by our methodology.\par\smallskip

Also, limitations in the rendering and OCR tools, including how rendering has changed over time and what was originally sent, may not be current. As a result, our ability to detect examples of concealment, which were only effective against older versions of mail clients, is hindered.\par\smallskip

Based on our findings, several areas for future research are indicated:
\begin{enumerate}
    \item \textbf{Concealment Analysis Metrics}: Compare the efficacy of difference metrics such as Kolmogorov Complexity or locality-sensitive hashing as an alternative to Jaccard distance when identifying occurrences of concealment depending on the concealment sub-type being employed.
    \item \textbf{Longitudinal Studies}: Conduct longitudinal studies to understand how content concealment strategies evolve over time and to assess the long-term effectiveness of different email filtering mechanisms, especially with the rise in prevalence of generative AI tools.
    \item \textbf{Advanced Filtering Techniques}: Investigating and developing advanced filtering techniques that can more effectively detect and mitigate content concealment sub-types, leveraging insights from this study. In particular filtering systems might detect suspicious use of CSS mechanisms.
\end{enumerate}

In conclusion, this study provides a detailed analysis of content concealment sub-types in emails, highlighting the challenges and techniques used by attackers. By identifying the specific sub-types and CSS tricks employed, we contribute to the understanding of how attackers evade email filters and offer insights for developing more effective detection methods.

%% file: 0Main.bbl

%% file: 9Appendix.tex
\input{examples/example-bulk2}
\input{examples/example-disrupt1}
\input{examples/example-insert1}
\input{latex-figures/concealment_examples}
\input{latex-figures/technique_examples}

%% file: examples/example-bulk2.tex
\begin{table}[h]
    \centering
    \textbf{Add Paragraph Email Example}\par\medskip
    \begin{tabular}{p{0.2\textwidth}p{0.7\textwidth}}
        \hline
        \textbf{Attribute} & \textbf{Details} \\ \hline
        \textbf{Email Year/Month} & 2007/07 \\ \hline
        \textbf{Filename} & 1185318861.15426\_491.txt \\ \hline
        \textbf{Rendered Email View} & 
        \includegraphics[width=0.7\textwidth]{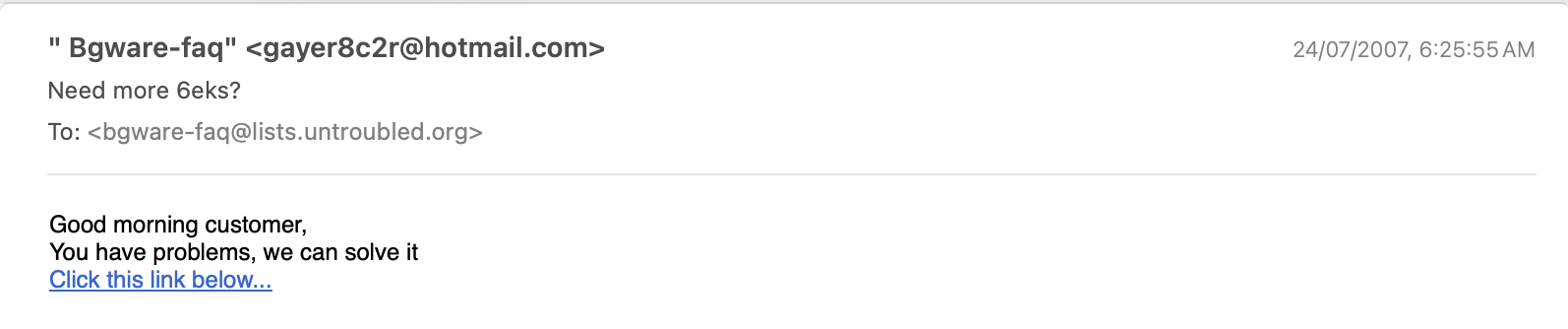} \\ \hline
        \textbf{Partial Email Source} & 
        \includegraphics[width=0.4\textwidth]{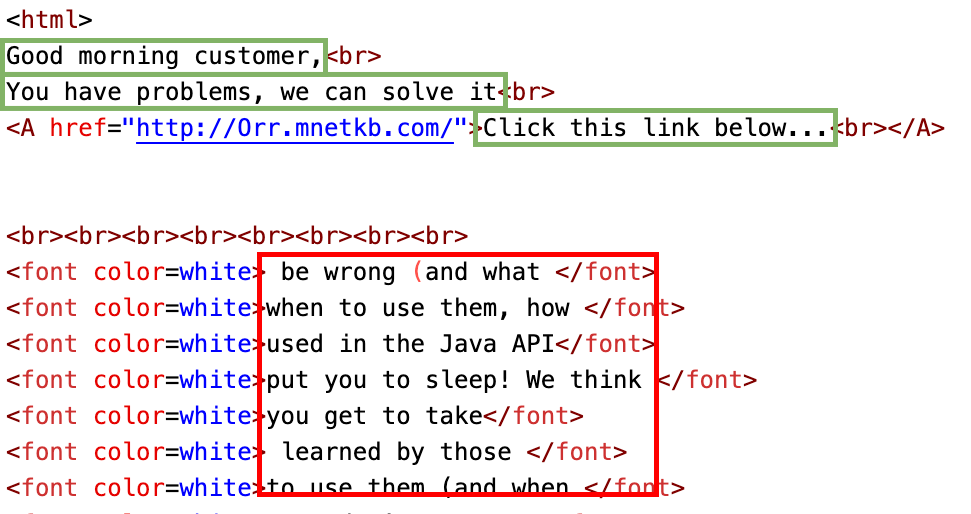} \\ \hline
        \textbf{Recipient View} & good morning customer you have problems we can solve it click this link below \\ \hline
        \textbf{Mail Filter View} & good morning customer you have problems we can solve it click this link below be wrong and what when to use them how used in the java api put you to sleep we think you get to take learned by those to use them and when you don t want to patterns look in to do instead you want what to expect a visually rich own with your co worker your time is too important used in the java api look in the wild to learn how those about inheritance might sounds how the factory secret language is so often misunderstood is so often misunderstood brain in a way that sticks or on the real relationship put you to sleep we think the trading spaces show you don t want to own with your co worker decorator is something from want to see how and adapter with head first when to use them how that you can hold your with design patterns of design patterns so the next time you re also want to learn \\ \hline
        \textbf{Sub-type} & Add Paragraph \\ \hline
        \textbf{CSS Trick(s)} & Font Colour\\ \hline
    \end{tabular}
    \caption{This figure contains the rendered email, partial email source, OCR output, plaintext, and a list of the concealment sub-types and CSS tricks used in the email. For the partial email source, the content displayed to the recipient is outlined in green, and the concealed content is outlined in red.}
    \label{fig:example_add_paragraph}
\end{table}

%% file: examples/example-disrupt1.tex
\begin{table}[H]
    \centering
    \textbf{Disrupt Word Email Example}\par\medskip
    \begin{tabular}{p{0.2\textwidth}p{0.7\textwidth}}
        \hline
        \textbf{Attribute} & \textbf{Details} \\ \hline
        \textbf{Email Year/Month} & 2004/08 \\ \hline
        \textbf{Filename} & 1091394472.23940\_30.txt \\ \hline
        \textbf{Rendered Email View} & 
        \includegraphics[width=0.7\textwidth]{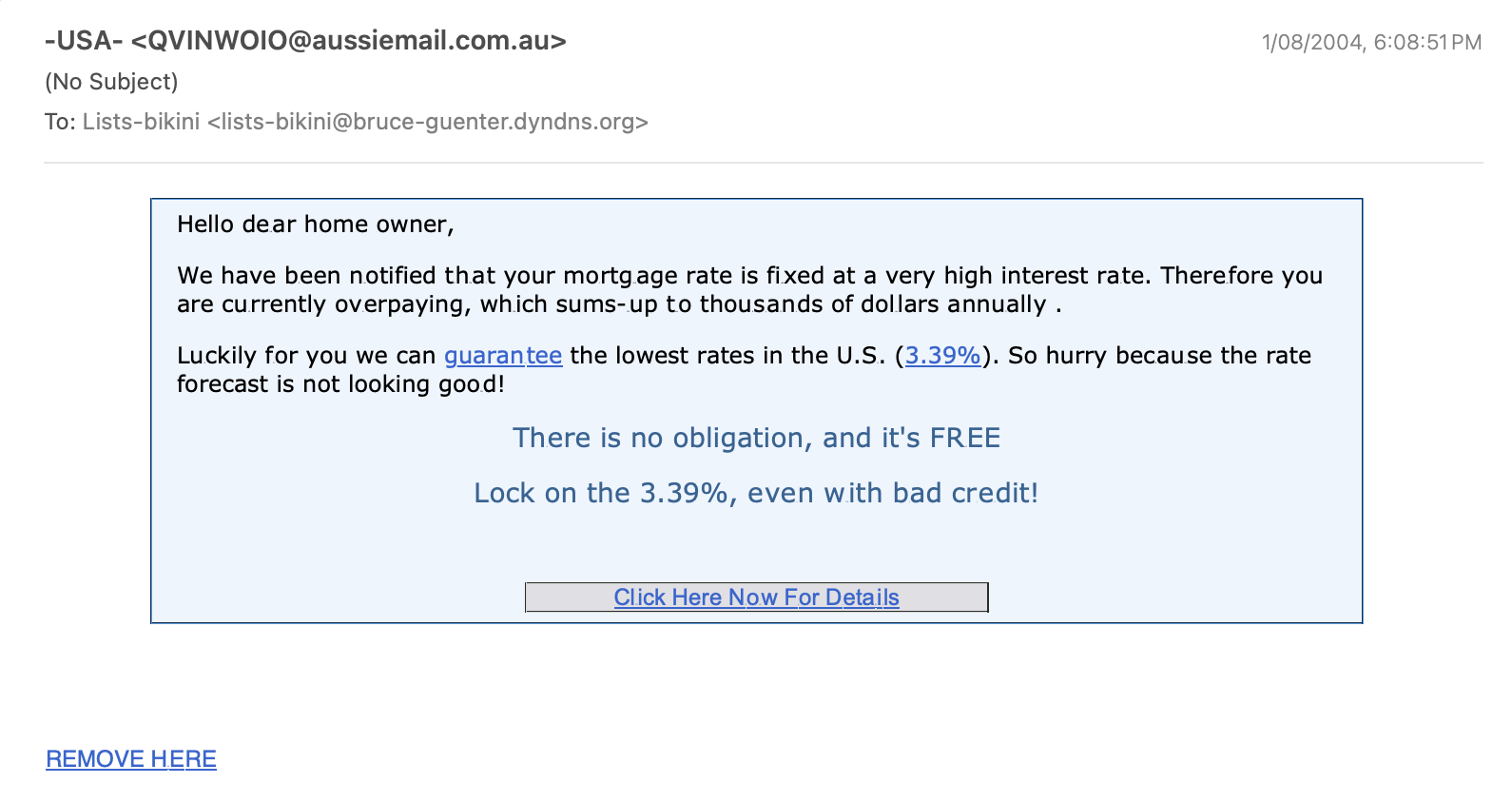} \\ \hline
        \textbf{Partial Email Source} & 
        \includegraphics[width=0.7\textwidth]{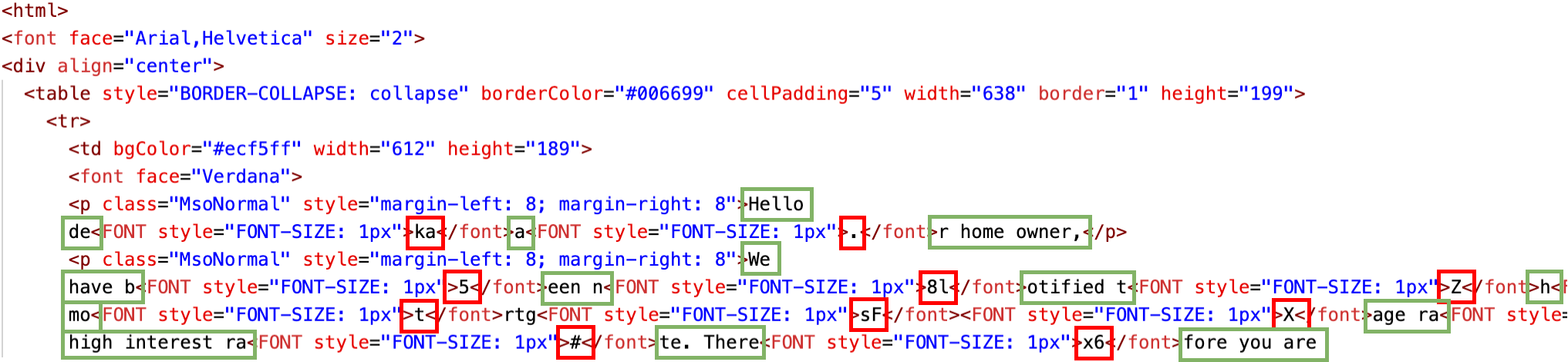} \\ \hline
        \textbf{Recipient View} & hello dear home owner we have been notified that your mortgage rate is fixed at a very high interest rate therefore you are currently overpaying which sums up to thousands of dollars annually luckily for you we can guarantee the lowest rates in the us 3 39 so hurry because the rate forecast is not looking good there is no obligation and it s free lock on the 3 39 even with bad credit click here now for details remove here \\ \hline
        \textbf{Mail Filter View} & hello dekaa r home owner we have b5een n8lotified tzhfma1t your motrtgsfxage ralte is fiqnxed at a very high interest ra te therex6fore you are cugtrrently ovtcerpaying whnkich sums 2kup tmvo tlhou6syarnzds of dolaxlars alnnu5all y lu3ckiily f5or you we can guarannjtee the lowest rates in the u sd 3 39 so hurry bec8aufmse the rate forecast is not looking goomrd thdere is no obligation and it s f rxdee l ock on the 3 39 e ve n wghith bjad cr e3dit clpsick her e ngo82w fkor dcetaqtials remove htoe re \\ \hline
        \textbf{Sub-type} & Disrupt Words \\ \hline
        \textbf{CSS Trick(s)} & Font Size\\ \hline
    \end{tabular}
    \caption{This figure contains the rendered email, partial email source, OCR output, plaintext, and a list of the concealment sub-types and CSS tricks used in the email. For the partial email source, the content displayed to the recipient is outlined in green, and the concealed content is outlined in red.}
    \label{fig:example_disrupt_word}
\end{table}

%% file: examples/example-insert1.tex
\begin{table}[H]
    \centering
    \textbf{Insert Word Email Example}\par\medskip
    \begin{tabular}{p{0.2\textwidth}p{0.7\textwidth}}
        \hline
        \textbf{Attribute} & \textbf{Details} \\ \hline
        \textbf{Email Year/Month} & 2005/07 \\ \hline
        \textbf{Filename} & 1121574039.27681\_5.txt \\ \hline
        \textbf{Rendered Email View} & 
        \includegraphics[width=0.7\textwidth]{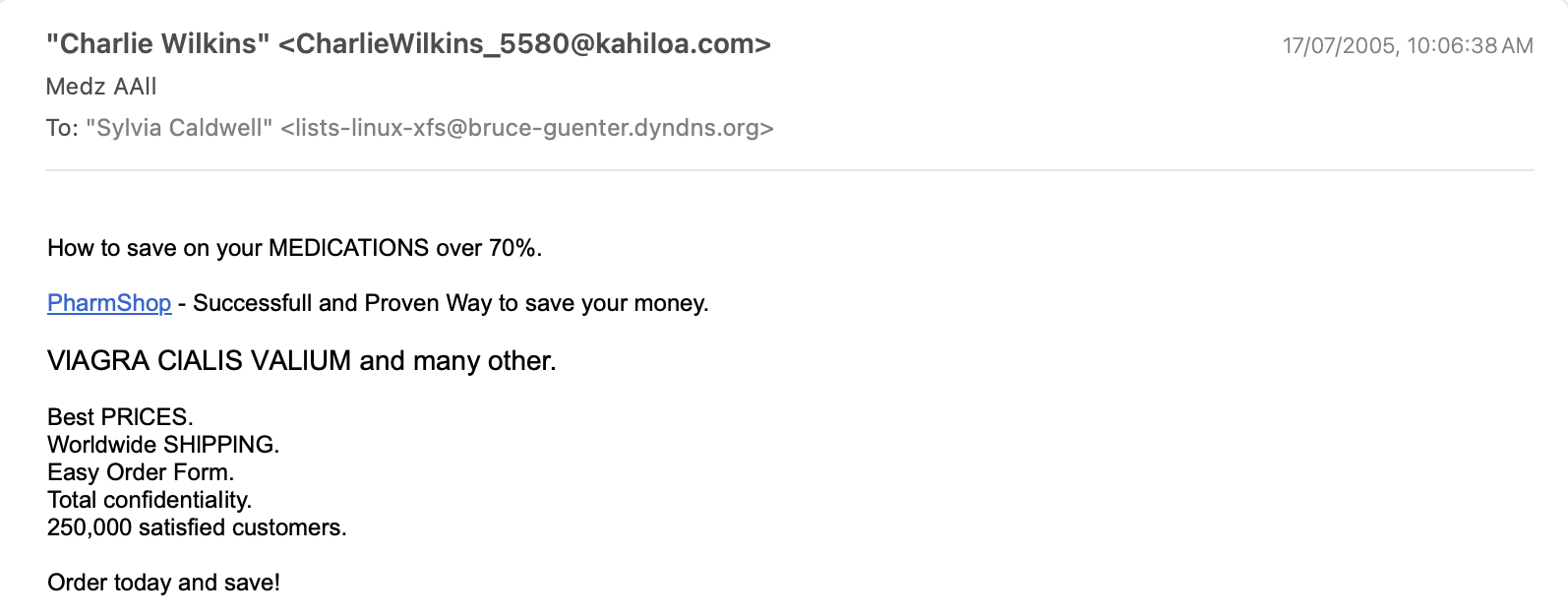} \\ \hline
        \textbf{Partial Email Source} & 
        \includegraphics[width=0.3\textwidth]{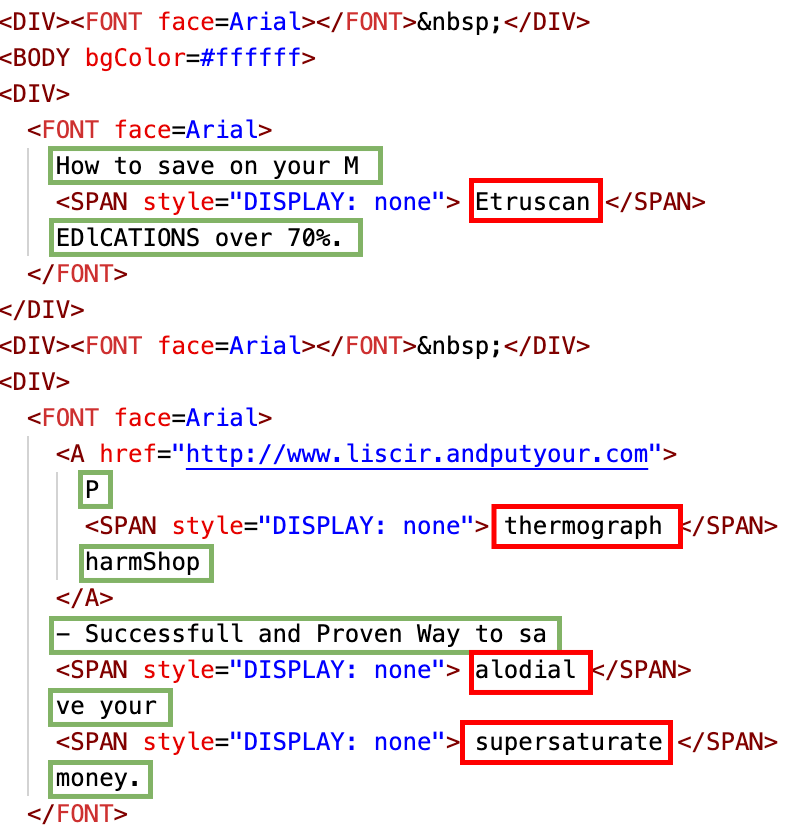} \\ \hline
        \textbf{Recipient View} & how to save on your medications over 70 pharmshop successfull and proven way to save your money viagra cialis valium and many other best prices worldwide shipping easy order form total confidentiality 250000 satisfied customers order today and save \\ \hline
        \textbf{Mail Filter View} & how to save on your m etruscan edlcations over 70 p thermograph harmshop successfull and proven way to sa alodial ve your supersaturate money playedout v resource ag a vilayet l intimacy lu selfhealing l inflorescence ra decipherable cl polychromatic is v galenic al respecting m and many other best p crotch rlces worl registrar dwide shlpplng ea depredate sy order form total confidenti perambulate aiity 250 000 satisf coralisland ied customers order to egregious day and save \\ \hline
        \textbf{Sub-type} & Insert Word \\ \hline
        \textbf{CSS Trick} & Position (Specifically display:none)\\ \hline
    \end{tabular}
    \caption{This figure contains the rendered email, partial email source, OCR output, plaintext, and a list of the concealment sub-types and CSS tricks used in the email. For the partial email source, the content displayed to the recipient is outlined in green, and the concealed content is outlined in red.}
    \label{fig:example_insert_word}
\end{table}

%% file: latex-figures/concealment_examples.tex
\renewcommand{\arraystretch}{1.3}
\begin{table}[ht]
\centering
    \textbf{Concealment Sub-type Examples}\par\medskip
\begin{tabular}{@{}rp{0.45\linewidth}p{0.41\linewidth}@{}}
\toprule
\textbf{Sub-type} & \textbf{Example} & \textbf{Explanation} \\ \hline
\textbf{Add Paragraph} & & \\
\textit{HTML Source} & {\textless}FONT color=\#fffffc size=2{\textgreater}prolate balfour rabid [...] pliant embroider{\textless}/FONT{\textgreater}[...]{\textless}font face=Arial color=\#000000{\textgreater}Each Order Includes[...] & The words ``prolate'' to ``embroider'' have font colour white with font size 2. \\\cline{2-3}
\textit{Mail Filter View}& prolate balfour rabid [...] pliant embroider Each Order Includes[...] & The mail filter sees additional words before the rendered text. \\\cline{2-3}
\textit{Recipient View}& Each Order Includes [...] & The concealed words are not visible to the recipient. \\\hline
\textbf{Disrupt Word} & & \\
\textit{HTML Source} & Pil{\textless}FONT style=``FONT-SIZE: 1px''{\textgreater}\#{\textless}/font{\textgreater}l{\textless}FONT style=``FONT-SIZE: 1px''{\textgreater}= /{\textless}/font{\textgreater}s & The word ``Pills'' has been interspersed with concealed punctuation characters using small font size. \\\cline{2-3}
\textit{Mail Filter View}& Pil\#l=/s & The mail filter does not see the same word as the recipient.\\\cline{2-3}
\textit{Recipient View}& Pills & The word ``Pills'' appears as normal to the recipient.\\\hline
\textbf{Insert Word} & & \\
\textit{HTML Source} & Dear {\textless}span style=``[...]''{\textgreater}lists-ezmlm{\textless}/span{\textgreater}, {\textless}i style=``font-size:11px; color:\#FFFAFA;''{\textgreater} never {\textless}br{\textgreater}[...]{\textless}a href=``[...]''{\textgreater}Forget about fear & The word ``never'' has been concealed before a linebreak by making the font colour similar to the background colour. \\\cline{2-3}
\textit{Mail Filter View}& Dear recipient never Forget about fear & The mail filter sees an additional word in the sentence. \\\cline{2-3}
\textit{Recipient View}& Dear recipient Forget about fear & The user sees a different version of the email. \\
\bottomrule
\end{tabular}
\caption{The three identified concealment sub-types are detailed in this table, including examples and explanations of the impact of the attack. These three examples have been modified from the sampled emails to suit space requirements, and extraneous content has been replaced with ``[...]''.}
\label{fig:concealment_quantities}
\end{table}

%% file: latex-figures/technique_examples.tex
\renewcommand{\arraystretch}{1.3}
\begin{table}[h]
\centering
    \textbf{CSS Trick Examples}\par\medskip
\begin{tabular}{@{}rp{0.45\linewidth}p{0.41\linewidth}@{}}
\toprule
\textbf{CSS Trick} & \textbf{Example} & \textbf{Explanation} \\ \hline
\textbf{Font Color} & & \\
\textit{HTML Source} & 
{\textless}A href=[...]{\textgreater}Cool shop in a one click{\textless}/A{\textgreater}
 [...] \par
{\textless}font color=white{\textgreater} In their native {\textless}/font{\textgreater}\par
{\textless}font color=white{\textgreater}(or worse, a flat tire), {\textless}/font{\textgreater}
& The text after the link is in a font color that matches the background, making it invisible to the recipient. \\\cline{2-3}
\textit{Mail Filter View}& Cool shop in a one click In their native (or worse, a flat tire), & The mail filter detects the text despite it being hidden in the rendered view. \\\cline{2-3}
\textit{Recipient View}& Cool shop in a one click & The concealed text is not visible to the recipient. \\\hline

\textbf{Font Size} & & \\
\textit{HTML Source} &
Please watch this one trade Tuesday and all week!{\textless}br{\textgreater}
[...]\par
{\textless}DIV{\textgreater}{\textless}FONT size=2{\textgreater}moth-eat sail twine{\textless}BR{\textgreater}\par
bracket capital south-southwesterly{\textless}BR{\textgreater}[...]{\textless}/FONT{\textgreater}
& The text after ``... all week!'' is rendered in a very small font size, making it difficult for the recipient to notice. \\\cline{2-3}
\textit{Mail Filter View}& Please watch this one trade Tuesday and all week! [...] moth-eat sale twine bracket capital south-westerly [...] & The mail filter can see the text despite its small size. \\\cline{2-3}
\textit{Recipient View}& Please watch this one trade Tuesday and all week! [...] & The text is very small and almost unnoticeable to the recipient, compared to normal text. \\\hline

\textbf{Text Position} & & \\
\textit{HTML Source} & 
{\textless}td align=``left''{\textgreater}Get the great di{\textless}span style=``FONT-SIZE: 2px; FLOAT: right; COLOR: white''{\textgreater} jzw {\textless}/span{\textgreater}scou{\textless}span style=``FONT-SIZE: 2px; FLOAT: right; COLOR: white''{\textgreater} kl {\textless}/span{\textgreater}nts on popular[...]
& The text ``jzw'', ``kl'' etc. is positioned on the right of the screen, in combination with small font size and white font colour, making it invisible to the recipient without impacting the positioning of the text it surrounds. \\\cline{2-3}
\textit{Mail Filter View}& Get the great di jzw scou kl nts on popular [...] & The mail filter sees the text despite its float position. \\\cline{2-3}
\textit{Recipient View}& Get the great discounts on popular [...] & The text is not visible to the recipient. \\\hline

\textbf{Table Manipulation} & & \\
\textit{HTML Source} & 
{\textless}td{\textgreater}Ac{\textless}br{\textgreater}Cl{\textless}br{\textgreater}De{\textless}br{\textgreater}[...]
{\textless}td{\textgreater}om{\textless}br{\textgreater}om{\textless}br{\textgreater}fl{\textless}br{\textgreater}[...]
{\textless}td{\textgreater}pl{\textless}br{\textgreater}id{\textless}br{\textgreater}uc{\textless}br{\textgreater}[...]
{\textless}td{\textgreater}ia{\textless}br{\textgreater}{\&}nbsp;{\&}nbsp;{\textless}br{\textgreater}an[...]
& Each table cell {\textless}td{\textgreater} contains multiple lines, each with a few characters. When combined other cells in the same row, these create visible and legible words, which might otherwise be considered spamlike. \\\cline{2-3}
\textit{Mail Filter View}& Ac Cl De[...] om om fl[...] pl id uc[...] is an[...] & The mail filter detects the text despite it being hidden. \\\cline{2-3}
\textit{Recipient View}& Acomplia\par
Clomid\par
Deflucan\par
[...]
 & The spam-like text is visible to the recipient. \\\hline

\textbf{Other} & & \\
\textit{HTML Source} & 
{\textless}STYLE{\textgreater}
DIV \{COLOR: \#FAFFFB\}
 DIV.b:first-letter \{COLOR: \#28ED2A\}
 DIV:first-letter \{FONT-SIZE: 300\%\}
 {\textless}/STYLE{\textgreater}\par
{\textless}DIV class{=}b{\textgreater}Seet!{\textless}/DIV{\textgreater}\par
{\textless}DIV class{=}b{\textgreater}Expired,{\textless}/DIV{\textgreater}\par
{\textless}DIV class{=}b{\textgreater}Lodgings,{\textless}/DIV{\textgreater}\par
{\textless}DIV class{=}b{\textgreater}Lake:{\textless}/DIV{\textgreater}
& As an example of the ``Other'' category, we chose one an example which does not fit any other category, but did not show up enough for its own category. In this example, each word uses its first letter to write the word ``SELL'' vertically in large, higher contrast letters. The rest of the content is concealed by the CSS color \#fafffb with low contrast against the background. \\\cline{2-3}
\textit{Mail Filter View}& Seet! Expired, Lodgings, Lake: & The email filter sees all of each word's text. \\\cline{2-3}
\textit{Recipient View}& S{\par}E{\par}L{\par}L & The word SELL is visible, written vertically. \\
\bottomrule
\end{tabular}
\caption{Examples of the CSS tricks identified and their potential effect on the mail filter. These examples have been modified from the sampled emails to suit space requirements, and extraneous content has been replaced with ``[...]''.}
\label{fig:css_techniques}
\end{table}